\documentclass[aps,twocolumn,floats,pre]{revtex4}
\usepackage{graphics,graphicx,epsfig}
\usepackage{amssymb}
\usepackage{epsf,epstopdf,wrapfig}

\newcommand{\sgn}{\mbox{sgn}}	
\begin{document}

\title{Thermodynamics of natural images}

\author{Greg J  Stephens,$^{a,}$\footnote{These authors contributed equally.}\, Thierry Mora,${}^{a,*}$ Gasper Tka\v{c}ik$^b$ and William Bialek$^{a,c}$}

\affiliation{$^a$Joseph Henry Laboratories of Physics,
$^a$Lewis--Sigler Institute for Integrative Genomics, and
$^c$Princeton Center for Theoretical Science,
Princeton University,
Princeton, New Jersey 08540\\
$^b$Department of Physics and Astronomy,
University of Pennsylvania,
Philadelphia, Pennsylvania 19104
}

\begin{abstract} 
The scale invariance of natural images suggests an analogy to the statistical mechanics of physical systems at a critical point.  Here we examine the distribution of pixels in small image patches and show how to construct the corresponding thermodynamics.  We find evidence for criticality in a diverging specific heat, which corresponds to large fluctuations in how `surprising' we find individual images, and in the quantitative form of the entropy vs.~energy. The energy landscape derived from our thermodynamic framework  identifies special image configurations that have intrinsic error correcting properties, and neurons which could detect these features have a strong resemblance to the cells found in primary visual cortex.
\end{abstract}

\maketitle

\section{Introduction}

From the familiar faces of our friends and family to objects of almost every size in environments of every type, the world that we see is full of structure.  Although this structure seems obvious when we look at the world, providing a precise mathematical description has proven more difficult.  One way to formulate this problem is to ask for a probability distribution of images such that, if we draw at random out of this distribution, the resulting images resemble those that we see in the natural environment.  Such a probabilistic or generative model would provide a rigorous basis for practical algorithms in image coding, processing and recognition \cite{mvBook}.   It is also reasonable to hypothesize that our brains have learned at least an approximation to  this probabilistic model, allowing us to form more efficient representations of the  visual world and to find efficient solutions of many seemingly difficult computational problems.  In this view, aspects of vision ranging from the responses of individual neurons to gestalt perceptual rules would be seen not as artifacts of the brain's circuitry but rather as matched to the statistical structure of the physical world \cite{barlow61,atick,simoncelli+olshausen_01,bialek_02}.

One statistical feature of natural images that provides a clue about the nature of the underlying probability distribution is scale invariance.  In particular, Field observed that the spatial patterns of image intensity from reasonably natural environments have power spectra that approximate $S_{I} \propto 1/k^2$, which is what one would expect from the hypothesis of scale invariance and simple dimensional analysis, and he suggested that this scaling behavior may have a direct connection to the distribution of receptive field parameters across neurons in visual cortex \cite{field87}.  The intuition that scale invariance is a strong constraint on the form of the probability distribution comes from statistical mechanics.  We recall that for systems in thermal equilibrium, the probability that we observe the system in state $\rm s$ is given by the Boltzmann distribution, $p_{\rm s} \propto \exp(-E_{\rm s}/T)$, where $E_s$ is the energy of the state and $T$ is the absolute temperature \cite{note1}.
For most physical systems  at generic values of the temperature and other parameters, correlations and power spectra are {\em not} scale invariant; rather there is some characteristic length $\xi$ that determines the distance beyond which structures approach statistical independence. Scale invariance emerges only when we tune the temperature to a special value $T_c$, the critical point which marks a second order phase transition between two different phases (liquid and gas, ferromagnet and paramagnet, ... ) \cite{cardy_96}.  The modern theory of critical phenomena teaches us that such scale invariance can occur while violating the naive expectations of dimensional analysis, so that power spectra can acquire ``anomalous dimensions,'' $S\propto 1/k^{2-\eta}$.  Further, scaling extends beyond low order statistics, so that the full probability distributions are predicted to be invariant (but non--Gaussian) under appropriate scaling transformations.  Both anomalous scaling and invariant non--Gaussian distributions for local features have been observed in an ensemble of natural scenes \cite{ruderman&bialek94,ruderman94}.

The analogy between scaling in natural images and the behavior of physical systems at their critical point point raises the question of whether there are analogs to the {\em thermodynamic} features of a critical point.  Can we, for example, generalize a given natural image ensemble to a family of ensembles indexed by a ``temperature,'' and show that there is something special (i.e., critical) about the temperature of the real ensemble?  If there is an analog of the diverging specific heat at $T_c$, what does this say about the nature of images?  What are the order parameters that characterize the underlying phase transition?  Here we report some preliminary results on these and related questions.

\begin{figure*}[t]
\includegraphics[width=0.9\linewidth]{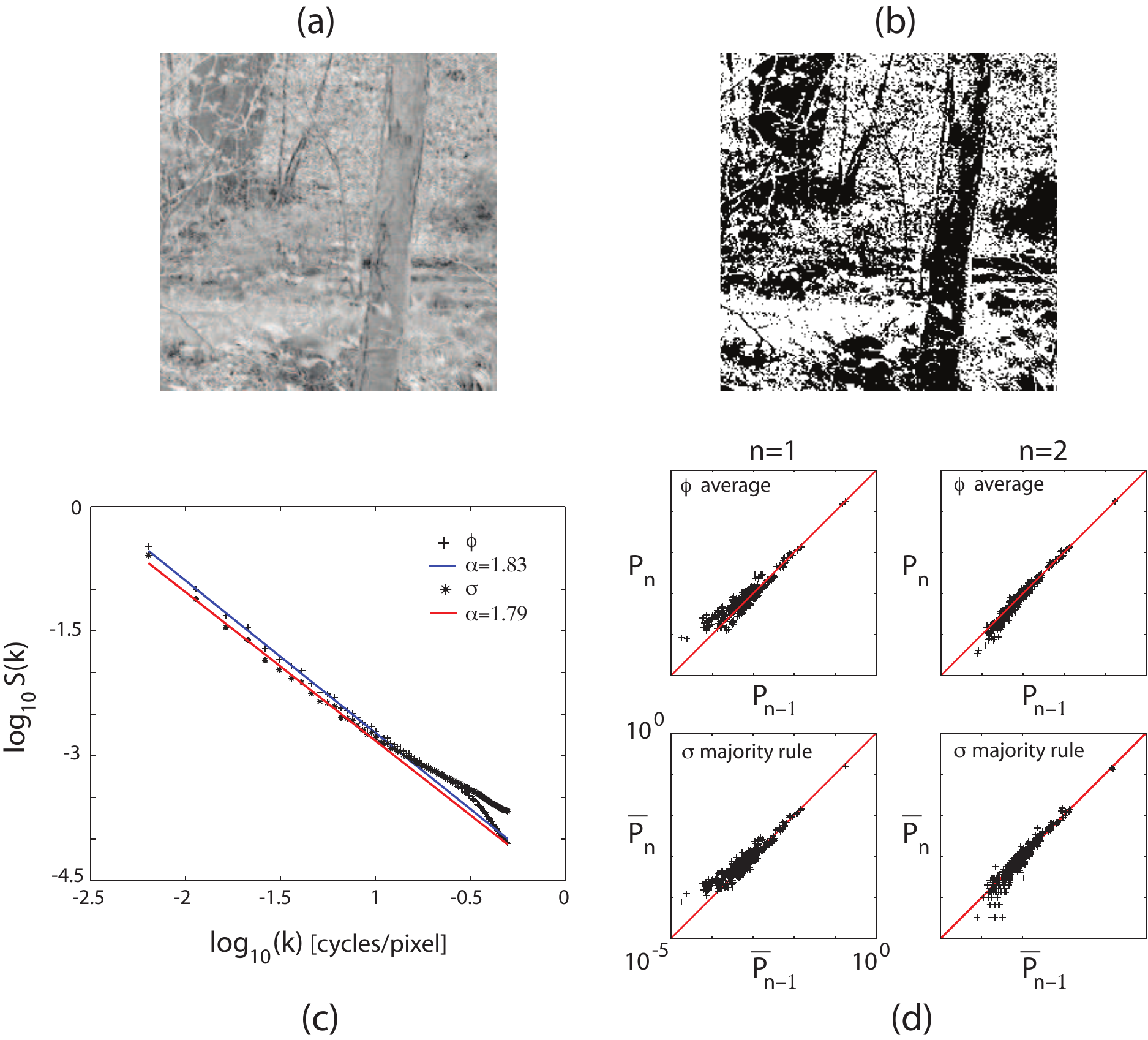}
\caption{Ensembles of natural images, and their quantized versions.  (a) An example image from the ensemble \cite{ruderman&bialek94}.  (b) The image from (a), after quantizing into two equally-poplulated levels.  Even when most intensity information is discarded the image retains substantial structure.  (c) Normalized power-spectra for gray-level and black and white images.  In both cases $S(k) \sim k^{-\alpha}$.  The small image size 
precludes a more accurate determination of the scaling exponent.  (d) The {\it full} distribution of black and white pixels in $3\times 3$ patches is
invariant  to block scaling.   The scaling is imposed either by block averaging the original light intensity and then quantizing (top) or by using the majority rule on quantized pixels (bottom).}
\label{fig:fig1}
\end{figure*}

\section{The image ensemble and scaling}

As an initial data set we returned to the image ensemble of Ref \cite{ruderman&bialek94}.  We focus here on the 45 images taken at lower spatial resolution, corresponding to $256\times 256$ pixel regions covering  $\sim 15^\circ\times15^\circ$ scenes in the woods of Hacklebarney State Park in New Jersey; an example is shown in Fig 1a.  Our path to the construction of a thermodynamics involves sampling  the distribution of images in small patches.  To make this problem manageable, we quantize the grey scale images into just two levels, with the quantization threshold chosen so that the numbers of black and white pixels are exactly equal over the ensemble.

It is important to verify that the rather harshly quantized images preserve interesting structures of the original scenes.  First, by inspection of Fig 1b we see that objects and even parts of objects (branches and leaves on the trees) are recognizable.  More quantitatively, if the original image is $\phi(\vec x )$, then we have constructed a discrete image $\sigma (\vec x) = \sgn[\phi(\vec x ) - \theta ]$, with $\theta$ chosen so that $\langle \sigma \rangle = 0$,  
where $\langle \cdots \rangle$ represents an average over the image ensemble.  Power spectra are defined by
\begin{equation}
\langle \phi(\vec x ) \phi(\vec x ')\rangle = \int {{d^2 k}\over{(2\pi )^2}}
S_\phi (\vec k ) e^{i\vec k \cdot (\vec x - \vec x ')} ,
\end{equation}
and similarly for $S_\sigma$.  In Fig 1c we show the spectra $S_\phi$ and $S_\sigma$, averaged over the orientation of the `momentum' vector $\vec k$ and normalized by the total variance.  We see that both spectra exhibit scaling, with very similar exponents.  As discussed in Ref \cite{ruderman&bialek94}, there is excess power at high frequencies because of aliasing, and more compelling evidence for scaling is obtained by combining these data with an ensemble of images from the same environment at higher angular resolution, so that the full range of $|\vec k|$ spans 2.5 decades.

In the quantized images, an $L\times L$ pixel region can take on $2^{L^2}$ possible states, and our data set provides $\sim 3\times 10^6$ samples of these states, although these are not independent.  Thus we can expect to provide a good sampling of the distribution of discretized image patches for $L=3$ or even $L=4$, since $2^{16}  \ll 3\times 10^6$, but $L=5$ is out of reach with this data set.  A direct estimate of the entropy shows that $S(4\times 4) = 11.154\pm 0.002\,{\rm bits}$, much less than the 16 bits we would obtained from random pixels; similarly, we find $S(3\times 3)=6.580\pm0.003<9\,{\rm bits}$. This quantifies our impression that a substantial amount of local structure is preserved in the discretized images.

We can also test for scaling more generally by asking how distribution of image states in $L\times L$ patches evolves when we  coarse--grain the images, and we can do this in two ways.  First, we can take the original grey scale images $\phi (\vec x )$ and create a new image such that the value of $\phi$ in each pixel of the new image is the average over a $2^{n}\times 2^{n}$ block of pixels in the original image, and then we can quantize these images.  When we look at $3\times 3$ patches in these filtered and quantized images, we again have $2^9$ possible states, and we call the distribution over these states $P_n$, where $P_0$ is the distribution obtained from the original images without any blocking.
In the same spirit (and following the original approach in Ref \cite{kadanoff}), we can take the quantized image $\sigma (\vec x )$ and directly create new quantized images by applying majority rule to the pixels in $3\times 3$ blocks, and this can be iterated; we'll call the resulting distributions of states in images patches $\bar P_n$, where again $\bar P_0 = P_0$ is what we obtain without blocking. Scale invariance is the claim that all the $P_n$ and $\bar P_n$ will be the same, independent of $n$, because the distribution of states is at a fixed point of this ``renormalization'' transformation \cite{cardy_96}.  In Fig 1d we test this prediction, showing that it is obeyed with good accuracy over four decades in probability.  There are more significant deviations in the first step of coarse--graining ($n=1$, at left in the figure), presumably because of the effects of aliasing noted above.  We emphasize that this test of scale invariance involves the full, joint distribution of image intensities in $3\times 3$ patches, and thus goes beyond checking the power law behavior of the spectrum (a second order moment) or the invariance of distributions of features (e.g., the outputs of local filters) evaluated at a single point. Similar results are obtained for $4\times 4$ patches.

\section{Temperature and specific heat}

Small patches of our discrete images are described by a set $\vec \sigma$ of binary variables.  Let us imagine that the distribution of these image patches is really the Boltzmann distribution for some physical system at temperature $T=1$, with some ``energy'' function $E(\vec \sigma )$ describing each possible patch,
\begin{equation}
P(\vec \sigma ) = {1\over {Z}} e^{- E(\vec \sigma )} .
\label{basic}
\end{equation}
Then, following the methods used in the analysis of dynamical systems \cite{feigenbaum,beck&schlogl}, we can define the distribution at any temperature $T$, since
\begin{eqnarray}\label{eq:pt}
P_T (\vec \sigma ) &\equiv& {1\over {Z(T)}} e^{-E(\vec \sigma )/T} = 
{1\over {{\cal Z} (T)}}[P(\vec \sigma )]^{1/T}\\
{\cal Z}(T) &=& \sum_{\vec\sigma} [P(\vec \sigma )]^{1/T} .
\end{eqnarray}
We can define an entropy at each value of the temperature,
$S(T) = -\sum_{\vec\sigma} P_T(\vec\sigma ) \log P_T (\vec\sigma )$, and then the usual thermodynamic relations tell us that the heat capacity is $C(T) = T \partial S(T)/\partial T$.  It is also useful to note that the heat capacity is proportional to the variance in energy or log probability, $C(T) = \langle [\delta E (\vec \sigma )]^2 \rangle_T/ T^2$, where $\langle \cdots \rangle_T$ denotes an average in the distribution $P_T(\vec \sigma )$.

\begin{figure*}[bt]
\includegraphics[width=\linewidth]{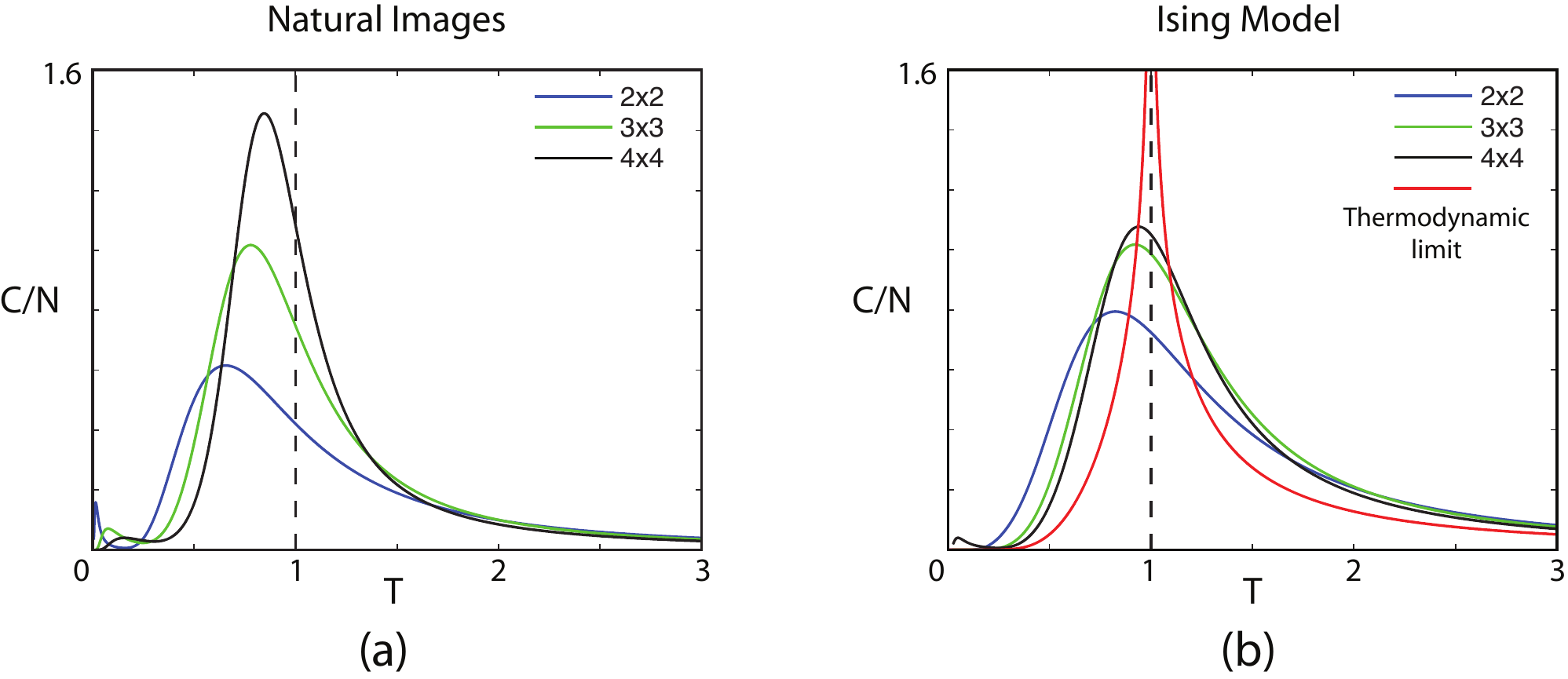}
\caption{Diverging specific heats of natural images.   (a) The specific heat $C/N=(T /N)\partial S(T) /\partial T$ constructed from natural images for $L\times L$ patches of linear dimension $L=2, 3, 4$  pixels.  Away from the natural operating temperature ($T=1$) the distribution is defined by Eq.~(\ref{eq:pt}). The peak in the specific heat near $T=1$ suggests that natural images are drawn from a critical ensemble. (b) As in (a) but
constructed  from Monte Carlo simulations of an Ising model with $T_c=1$; also shown is the exact behavior in the thermodynamic limit \cite{onsager44}.  Even in small patches we see hints of the underlying  critical behavior.}
\label{fig:fig2}
\end{figure*}

In a system with a critical point, the specific heat should diverge at $T=T_c$.  Of course this is true only in the thermodynamic limit of large systems, corresponding here to image patches containing many pixels.  Can we see precursors of this divergence in the small patches that we can actually sample?  Figure 2a shows the specific heat for $2\times 2$, $3\times 3$ and $4\times 4$ patches in our image ensemble, calculated directly from our sampling of the distributions $P(\vec \sigma )$.  We see that, even when we normalize by the number of pixels $N=L^2$ in each patch (since we expect that the heat capacity is extensive), looking at larger patches reveals a larger specific heat with a clear peak as a function of temperature, and this peak is shifting toward $T=1$.

To calibrate our intuition about the specific heat estimated from small patches, we have done precisely analogous computations on the nearest neighbor ferromagnetic Ising model in two dimensions, defined by $E(\vec \sigma ) = -J \sum_{({\rm ij})} \sigma_{\rm i} \sigma_{\rm j}$, where  $\sum_{({\rm ij})}$ denotes a sum over neighboring pairs of pixels. Monte Carlo simulations of this model generate binary ``images,'' and so many of the practical sampling questions are very similar to those in our problem.  We see in Fig 2b that the specific heat again shows a peak which grows and moves toward the true critical temperature $T_c =1$ as we look at larger patches.  Quantitatively the behavior is actually less dramatic than in the images, perhaps because the divergence of the specific heat in the thermodynamic limit is very gentle (logarithmic). Although this Ising spin system certainly is much simpler than the ensemble of images, comparison of Fig 2a and 2b supports the idea that we what we see in the images is consistent with an underlying divergence of the specific heat at a critical temperature close the to real temperature $T=1$.

\section{Entropy vs. energy and Zipf's law}

A complementary perspective on thermodynamics is the microcanonical ensemble, corresponding to  fixed energy rather than fixed temperature.  The end result of the discussion will be an attempt to measure, for our image ensemble, the entropy as a function of energy.  We begin with some standard results on how thermodynamic quantities are encoded in the plot of entropy vs. energy and on how we can identify a critical point in this plot.

\begin{figure*}[bt]
\includegraphics[width=\linewidth]{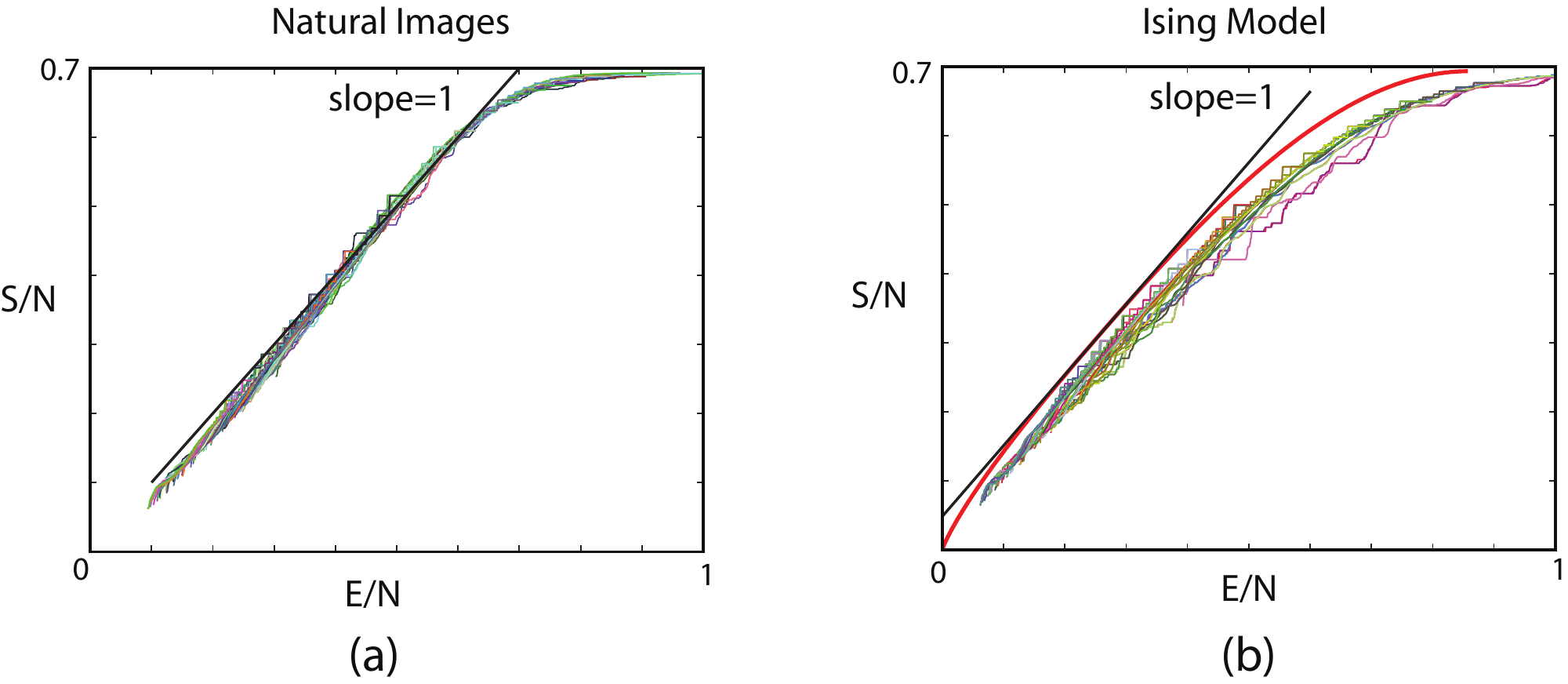}
\caption{Indication of critical behaviour in the plot of entropy vs. energy. (a) The results for rectangular image patches of size 8 to 50 pixels. (b) Results for Monte-Carlo simulations of the 2-D Ising model. The smooth red curve in the exact thermodynamic limit \cite{onsager44}.}
\label{fig:fig3}
\end{figure*} 

All thermodynamic quantities can be recovered from the partition function,
\begin{equation}
Z(T) = \sum_{\vec \sigma} e^{-E(\vec\sigma ) /T} .
\end{equation}
We can rewrite this sum by grouping together all states that have the same energy, 
\begin{widetext}
\begin{equation}
Z(T) = \sum_{\vec \sigma} e^{-E(\vec\sigma ) /T} 
= \int dE \left[ \sum_{\vec \sigma} \delta \left( E - E(\vec\sigma )\right) \right] e^{-E/T} = \int dE\,\rho(E) e^{-E/T} ,
\end{equation}
\end{widetext}
which defines the density of states $\rho(E)$.  For a large system (patches with many pixels), the density of states becomes a smooth function, and we can define an entropy $S(E)$ at fixed energy as the log of the number of states in a narrow range of energies, so that $\rho(E) = (1/\Delta ) e^{S(E)}$.  Then the partition function is 
\begin{equation}
Z(T) = {1\over \Delta} \int dE \,e^{S(E) - E/T} .
\end{equation}
Further, both the energy and entropy are extensive variables which should be proportional to the size of the system, $N$; here $N$ will be the number of pixels in a patch. Then we define $\epsilon = E/N$ and $s(\epsilon ) = S(E = N\epsilon )/N$, and the partition function becomes
\begin{equation}
Z(T) = {N\over \Delta} \int d\epsilon \,e^{N[s(\epsilon ) - \epsilon/T]} .
\end{equation}
Now it is clear that, as $N$ becomes large, the integral will be dominated by the point where exponent is maximal, that is an energy such that $ds(\epsilon )/d\epsilon = dS(E)/dE = 1/T$.  This connects the (microcanonical) description at fixed energy with the (canonical) description at fixed temperature.  In addition, one can show that the specific heat is (inversely) related to the second derivative of the entropy, 
\begin{equation}
C = {N\over {T^2}}\left[ - {{d^2 s(\epsilon )}\over{d\epsilon^2}}\right]^{-1} .
\end{equation}
In this language, the divergence of the specific heat occurs where the second derivative of the entropy vs. energy vanishes. This is the hallmark of a second order phase transition.

It is important to note that we can define $E$ for every state that we observe simply as the log of the probability, from Eq (\ref{basic}), 
$E(\vec\sigma ) = -\ln P(\vec\sigma ) + c$, where $c$ defines the (arbitrary) zero of energy, which we choose so that the most probable state has zero energy.  While it is tricky to measure the density of states or distribution of energies, it is easy to define the cumulative distribution, 
\begin{equation}
{\cal N}(E) = \int_0^E dE'\,\rho(E' ) ,
\end{equation}
which just counts the number of possible image patches for which the observed log probability is greater than $-E+c$.
If $S(E)$ is increasing, then this integral is dominated by the behavior near its upper limit, so that
\begin{eqnarray}
{\cal N}(E) &=& {N\over \Delta } \int_0^{E/N} d\epsilon \,e^{Ns(\epsilon )} \\
&\approx&  {N\over \Delta} \left[ N{{ds(\epsilon )}\over{d\epsilon}}\right]^{-1} e^{Ns(\epsilon = E/N)}\\
\Rightarrow s(\epsilon ) &=& {1\over N} \ln {\cal N}(E =N\epsilon) + {{\ln(T/\Delta )}\over N} .
\end{eqnarray}
Notice that the second term in this equation vanishes for large $N$, and so we approximate the entropy per pixel as a function of energy per pixel by the first term. 

The results of the previous paragraph imply  that if we just count the number of possible image patches with probability greater than a certain level, then we can construct the entropy vs. energy and hence derive all other thermodynamic functions.  This is what we do in Fig. 3a for our image ensemble, using rectangular $L\times L'$ patches  of size from 8 pixels up to 50 pixels; in Fig 3b we do the same thing for our Monte Carlo simulations of the Ising model.  As a practical matter it is important to note that sampling problems are less serious at low energies (states with high probability), so we expect that even if we look at regions where we can't sample the whole distribution we will get the correct low energy behavior.

We first note  that the results on image patches of different sizes are remarkably consistent with one another, suggesting that we are seeing signs of the thermodynamic limit despite the small size of the regions that we can explore fully.  Next, since the real image ensemble is at $T=1$, we want to pick out the energy at which $dS/dE = 1$; this seems very difficult, since the plot is very nearly a straight line with unit slope.
But we know what this means:  if the point where $dS/dE = 1$ is also a place where $d^2 S/dE^2 =0$, then $T=1$ is a critical point.  
Thus, {\em the fact that $S/N$ vs. $E/N$ is very nearly a straight line of unit slope is direct evidence that the ensemble of natural images is at criticality.}

\begin{figure*}
\includegraphics[width=\linewidth]{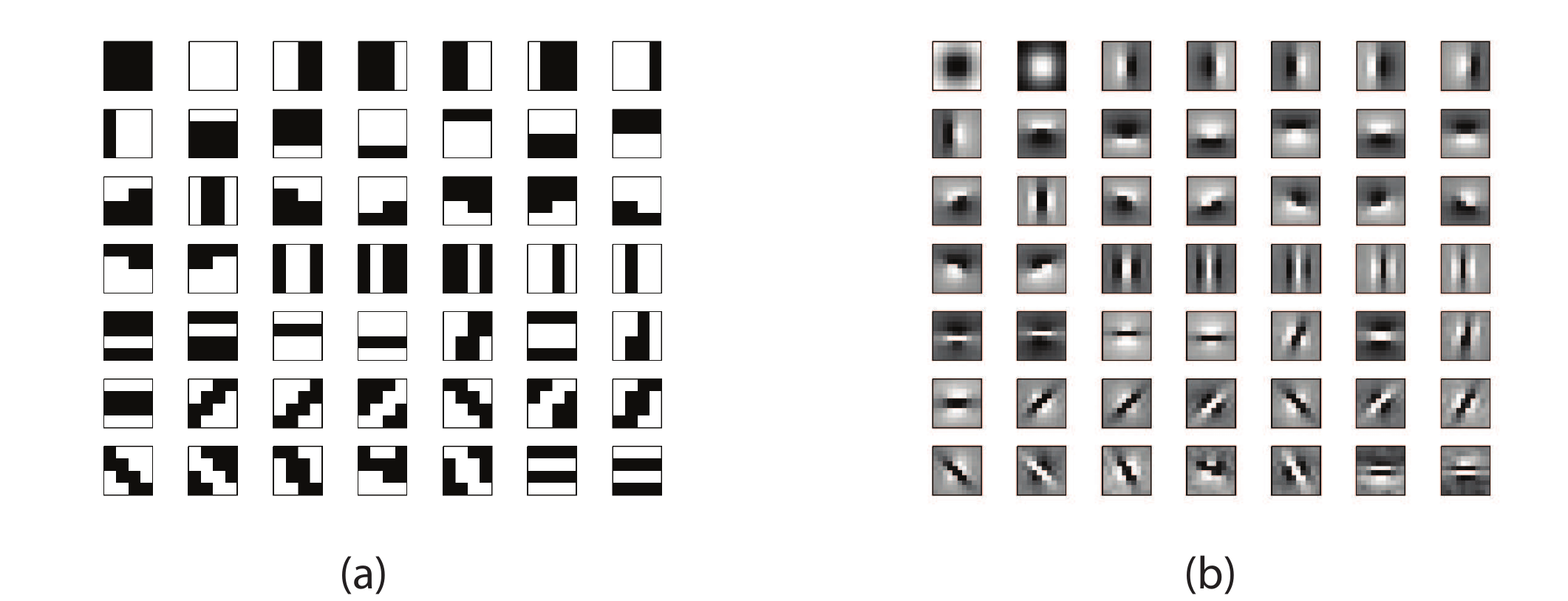}
\caption{Locally stable states.  (a) The 49 most probable patches such that all single pixel inversions {\it decrease}
their probability in the ensemble.  Even in small ($4\times4$) patches these locally stable states are interpretable as lines and edges at all positions and orientations. (b) The average surrounding $10\times 10$ light-intensity images leading to these metastable states. These images resemble the average images that trigger responses of neurons in primary visual cortex.}
\label{fig:fig4}
\end{figure*} 

It is interesting that this approach to the thermodynamics of images is connected to Zipf's law \cite{zipf}.  We recall that Zipf estimated the probability distribution from which words are drawn in an English text, and argued that if we put the words in rank order ($r=1$ is the most common word), then $p_r \propto 1/r$ up to some maximum rank $r= N_w$ corresponding to the number of different words $N_w$ used in the text.  Subsequently, other authors have considered generalized Zipf--like distributions \cite{newman}, $p_r \propto 1/r^\alpha$, and there has been much discussion about the meaning of these relationships. Suppose we identify the Zipf--like distribution $p_r = A/r^\alpha$ with a Boltzmann distribution at $T=1$, $p_r = (1/Z)e^{-E_r}$.  Then the energy of the state at rank $r$ is 
$E_r = \alpha \ln r - \ln(AZ)$. In the limit of a large system with many possible states, we can approximate the density of states by realizing the variable $r$ has a uniform distribution, and hence $\rho(E) \approx | dE_r /dr|^{-1}$; this gives $\rho(E) = r/\alpha$.  But we also have $r = (AZ)^{1/\alpha} e^{E_r/\alpha}$, so we find 
\begin{equation}
\rho(E) = {1\over\alpha} (AZ)^{1/\alpha} e^{E/\alpha} \Rightarrow S_{\rm Zipf} (E) = E/\alpha + {\rm constant} .
\end{equation}
Thus a generalized Zipf's law is equivalent to an entropy that is (exactly!)  linear in the energy.  The original Zipf's law ($\alpha =1$) corresponds to a unit slope,   as we have found for image patches.  Further, we have seen that this simple linear relation corresponds exactly to what we find for a thermodynamic system at a critical point \cite{note2}.

Why does it matter that the ensemble of natural images is at a critical point?  The signature of criticality is the divergence of the specific heat, and the specific heat is the variance in the energy, which is the log probability. Thus, being at a critical point means that the log probability has an enormously broad distribution, with a formally divergent second moment even once we normalize by the number of pixels.  One consequence is that the approach toward typicality in the sense of information theory \cite{shannon48} will be much slower than one would find away from the critical point, which may be related to difficulties in compressing large natural images, or even in estimating their entropy (see, for example, Ref \cite{chandler&field07}).  The large variance in log probability also means that there are large fluctuations in how surprised we should be by any given scene or segment of a scene, which perhaps quantifies our common experience.

\section{The energy landscape}

Critical points mark the transition between phases characterized by different forms of order:  liquid vs. gas, ferromagnet vs. paramagnet, and so on.  What is the ordering that would emerge if somehow the distribution of natural images could be ``cooled'' from $T=1$ down toward $T=0$?  This ultimately is a question about the nature of the image patches that correspond to the low energy states.  Certainly the lowest energy states of small patches are solid black or white blocks, as in a ferromagnet where all the spins can align up or down, and these states will dominate at $T=0$.  But, searching through all $4\times 4$ patches, we find $\sim 100$ states that are local minima of the energy, in the sense that flipping any single pixel from black to white (or vice versa) results in increased energy or reduced probability.  In Figure 4a we show 49 of these states, ordered in decreasing probability.  We see that many of these states are interpretable, for example as edges between dark and light regions, and that much of the multiplicity arises from the different ways of realizing these patterns (e.g., the six possible cases of a single vertical edge).  We can think of these local minima in the energy landscape \cite{spin_glass} as being like the attractors in the Hopfield model of neural networks \cite{hopfield82}, or like the code words in statistical mechanics approaches to error--correcting codes \cite{sourlas}.  Usually we think of error--correcting coding as a construct, but here it seems that the signals which the world presents to us have some intrinsic error--correcting properties.

Although there are many reasons why edges may be important for vision, it is interesting to take seriously the idea that such image features acquire their importance because of their intrinsic properties of error correction, as if these are the signals that the world is ``trying'' to send us in the most fault tolerant fashion.  If this is the case, then the visual system might build feature detecting neurons which serve to identify the basins of attraction defined by these local minima in energy.  If such cells respond only when the original grey scale image corresponds to a discrete image within a particular basin of attraction, then it is easy to compute the response--triggered average within our natural image ensemble, with the results shown in Fig 4b.  These results have a strong resemblance to the spike--triggered average responses of neurons in visual cortex to natural scenes \cite{sta}.  Interestingly, if we look more closely at the problem of identifying the basin of attraction, we find that perceptron--like models based on filtering through a single receptive field do rather poorly.  Thus, if visual cortex really builds a representation of the world based on the identification of these local minima in the energy landscape, the computations involved necessarily involve nonlinear combinations of multiple filters, as observed \cite{rust+al_05}.  Much remains to be done to see if this really is a path to a theory of these more complex neural responses.

\begin{acknowledgments}
We thank D Chigirev, SE Palmer \& E Schneidman for helpful discussions, and DL Ruderman for recovering the original data from Ref \cite{ruderman&bialek94}.  This work was supported in part by National Science Foundation Grants IIS--0613435, IBN--0344678, and PHY--0650617, by National Institutes of Health Grant T32 MH065214, by the Human Frontier Science Program, and by  the Swartz Foundation. 
\end{acknowledgments}


\begin{thebibliography}{99}
\bibitem{note1}
We simplify our notation by measuring temperature and energy in the same units.
\bibitem{note2}
Interestingly, if $\alpha\neq 1$ then there is no correspondence with thermodynamics.
\end{thebibliography}
\end{document}